\documentclass{ws-procs9x6}

\newcommand{\be}{\begin{equation}}
\newcommand{\ee}{\end{equation}}
\newcommand{\ds}{{\sf DarkSUSY}}

\begin{document}

\title{GAMMA RAYS FROM DARK MATTER}

\author{T. BRINGMANN}

\address{II. Institute for Theoretical Physics, University of Hamburg,\\
Luruper Chaussee 149, D-22761 Hamburg, Germany\\
E-mail: torsten.bringmann@desy.de\\
http://unith.desy.de}

\begin{abstract}

A leading hypothesis for the nature of the elusive dark matter are thermally produced, weakly interacting massive particles  that arise in many theories beyond the standard model of particle physics. Their self-annihilation in astrophysical regions of high  density provides a potential means of indirectly detecting dark matter  through the annihilation products, which nicely complements direct and collider searches.
Here, I review the case of gamma rays which are particularly promising in this respect: distinct and unambiguous spectral signatures would not only allow a clear discrimination from astrophysical backgrounds but also to extract important properties of the dark matter particles;  powerful observational facilities like the Fermi Gamma-ray Space Telescope or upcoming large, ground-based Cherenkov telescope arrays will be able to probe a considerable part of the underlying, e.g. supersymmetric, parameter space. I conclude with a more detailed comparison of indirect and direct dark matter searches, showing that these two approaches are, indeed, complementary.

\end{abstract}


\bodymatter

\section{Introduction}

The existence of a sizable non-baryonic, cold dark matter (DM) component  in the universe is supported by a wealth of observations on distance scales that range from individual galaxies over clusters of galaxies all the way out to fluctuations of the cosmic microwave background, spanning thus a range from tens of kpc to several Gpc. 
The \emph{amount} of DM,  $\Omega_\chi=0.227\pm0.014$ according to the most recent estimates\cite{wmap7} $\!\!$, can now be determined  much more precisely than in the times of Zwicky who  who  first formulated the DM problem almost 80 years ago \cite{zwicky};
as of today, however, the \emph{nature} of DM still remains an open question\footnote{
Actually, the situation has recently become quite interesting as both direct\cite{direct_DM} and cosmic ray\cite{cosmicray_DM} experiments have reported data that could be interpreted in terms of DM; at present, however, no compelling evidence for the DM and against competing, conventional explanations has been presented.}. 

A theoretically particularly appealing solution is a class of DM candidates that go under the name of Weakly Interacting Massive Particles (WIMPs): new effects related to electroweak symmetry breaking (currently being searched for at the CERN LHC) generically predict such particles which would be produced as thermal relics with the right density today \cite{DM_reviews}. While the lightest supersymmetric neutralino is often taken as a standard template for such a WIMP, alternative candidates like the lightest Kaluza-Klein photon in theories with universal extra dimensions \cite{Hooper:2007qk} or the lightest $T$-odd particle in little Higgs models \cite{Birkedal:2006fz} provide interesting alternatives.
Apart from being well-motivated from particle physics, WIMPs have the advantage of being, at least in principle, detectable by means other than gravitational: at colliders (where the main signature would be missing transverse energy), in direct detection experiments (where one hopes to measure the recoil energy of WIMPs off the nuclei of terrestrial  detectors) or indirectly through the observation of WIMP annihilation products in cosmic rays.

Among possible messengers for indirect DM searches, there are several advantages connected to gamma rays: usually produced at rather high rates, they propagate unhindered through the galactic halo; this means that they point directly back to the source and no assumptions about the diffusive halo are necessary like in the case of charged particles. What is maybe even more important is that they provide very clear and distinct spectral signatures to look for -- a point which will now be taken up in more detail.

\section{Spectral signatures}

The differential gamma-ray flux from a source with DM density $\rho$ in the direction $\psi$ is given by
\be
\label{flux}
	\frac{d\Phi_{\gamma}}{dE_\gamma} (E_\gamma,\psi) = {\frac{\langle\sigma v\rangle_\mathrm{ann}}{4\pi m_{\chi}^2} \sum_f B_f\frac{dN_\gamma^{f}}{dE_\gamma}}\times\frac{1}{2}{\int_{\Delta\psi}\frac{d\Omega}{\Delta\psi}\int_\mathrm{l.o.s} d\ell(\psi) \rho^2(\mathbf{r})}\,,
\ee
where $\langle\sigma v\rangle_\mathrm{ann}$ is the total annihilation cross section, $m_\chi$ the mass of the DM particle, $B_f$ the branching ratio into channel $f$ and  $N_\gamma^{f}$ the number of resulting photons; these quantities depend only on the underlying particle physics model and can thus be determined to a rather good accuracy for any given DM model. The right part of the above expression is a measure for the number of DM particle pairs along the line of sight of a detector with an opening angle $\Delta\psi$; since it depends strongly on the largely unknown distribution of DM, there is usually a considerable uncertainty connected to the overall \emph{normalization} of the expected annihilation flux.\footnote{
The gamma-ray flux only factorizes as indicated in Eq.~(\ref{flux}) if $\langle\sigma v\rangle_\mathrm{ann}$ is largely independent of the relative velocity $v$ of the annihilating particles; while this is often the case (valid to a very high accuracy for $s$-wave annihilations, in particular), there are situations (e.g.~pure $p$-wave or Sommerfeld enhanced annihilations) where this is not the case and one has to perform the angular average and line integral over the full expression.}

In order to claim a DM detection in gamma rays, i.e.~a convincing discrimination of the signal from more conventional astrophysical sources, it is thus important to rely on clear \emph{spectral signatures}.
In general, one can distinguish four different types of contributions to the gamma-ray spectrum: 

\begin{enumerate}
\item At tree-level, WIMPs
annihilate into pairs of quarks, leptons, Higgs and weak gauge bosons. \emph{Secondary photons} are produced in the hadronization and
further decay of these primary annihilation products, mainly through $\pi_0\rightarrow\gamma\gamma$. The result is a featureless spectrum with a rather soft cutoff at $m_\chi$, almost indistinguishable for the various 
possible annihilation
channels (with the exception of $\tau$-lepton final states)\cite{Bertone:2006kr}.

\item Whenever charged annihilation products are present, additional \emph{internal bremsstrahlung} (IB) photons appear at $\mathcal{O}\left(\alpha_\mathrm{em}\right)$. They generically dominate
at the highest energies that are kinematically accessible, and thereby
add pronounced signatures to the spectrum; viz.~a very sharp cutoff at $m_\chi$ and bump-like features at
slightly smaller energies.\cite{FSR,IB_SUSY} More recently, it has been pointed out that also electroweak corrections, with a final state $W$ or $Z$ boson instead of a photon, can give sizable corrections that could visibly effect the spectrum\cite{EW_corr}. 

\item Necessarily loop-suppressed, and thus only at $\mathcal{O}\left(\alpha_\mathrm{em}^2\right)$, 
\emph{monochromatic} $\gamma$
\emph{lines} result from the  annihilation of DM particles into two-body final states containing a photon \cite{lines}.
While in principle providing a striking
signature, these processes are usually subdominant and thus not actually visible  when taking into account realistic detector  resolutions \cite{IB_SUSY,KK_lines};
examples of particularly strong line signals, however, exist \cite{idm}.

\item In models with large branching fractions into $e^+/e^-$ pairs, finally, there is also a contribution from \emph{inverse Compton} scattering of highly energetic $e^\pm$ on starlight and the cosmic microwave background. It only appears at energies considerably below $m_\chi$, however, and does not result in pronounced spectral signatures  like in the previous two cases\cite{IC_DM}.

\end{enumerate}

\begin{figure}[t]
\vspace*{-0.1cm}
\begin{minipage}[t]{0.50\textwidth}
\centering
\includegraphics[width=\textwidth]{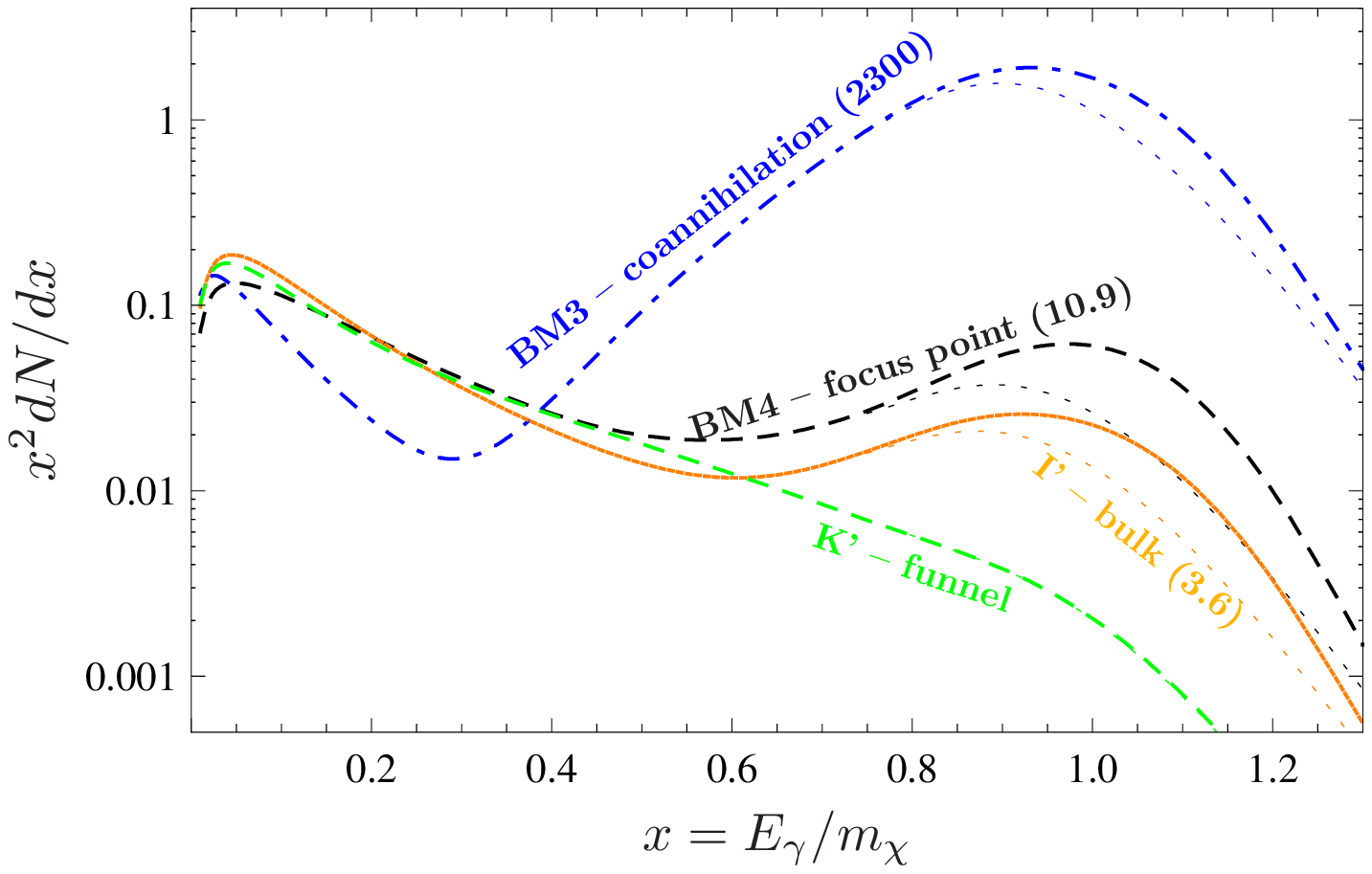}\\
\end{minipage}
\begin{minipage}[t]{0.01\textwidth}
\end{minipage}
\begin{minipage}[t]{0.48\textwidth}
\centering  
\vspace*{-3.69cm}
\includegraphics[width=\textwidth]{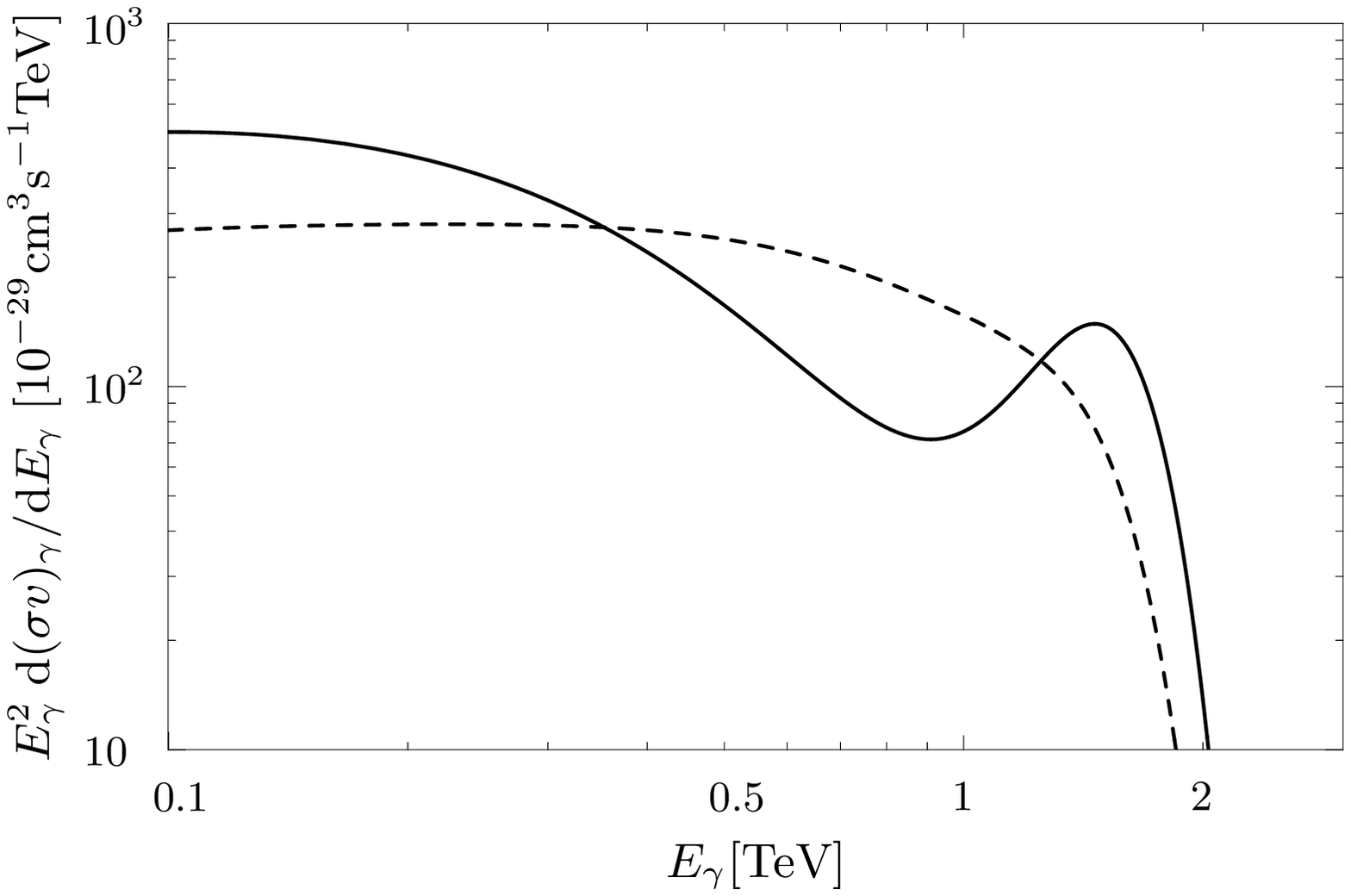}\\
\end{minipage}
\caption{
This figure shows typical DM annihilation spectra: in the left panel, four benchmark models representative of the four cosmologically relevant regions in the mSUGRA parameter space are plotted, as roughly seen by a detector with an energy resolution of 10\% (from Ref.~\refcite{Bringmann:2008jn}). For each of these models, the  number of IB over secondary photons (at energies $E_\gamma>0.6\,m_\chi$)  is indicated in parenthesis and the dotted line shows the same spectrum without including the line signal. 
The right panel (from Ref.~\refcite{Bergstrom:2006hk}) compares the spectra of Wino (solid line) and Kaluza-Klein DM (dashed line), assuming an energy resolution of $15\%$ and, for the sake of comparison, a common mass of $m_\chi=1.5\,$TeV.
\label{fig_spectra}}
\end{figure}

The distinct spectral features connected to IB even remain visible when taking into account realistic energy resolutions for the detector. As an example, consider the case of minimal supergravity (mSUGRA) where one can single out four regions in the underlying parameter space that give the correct DM relic density (see, e.g., Ref.~\refcite{Battaglia:2003ab} for a discussion). The left panel of Fig.~\ref{fig_spectra} shows the type of annihilation spectra typically encountered in these cases, smeared with a Gaussian to 
simulate the effect of an energy resolution of $10\%$. 
Clearly, one could use these spectra -- even in their smoothed versions --  to discriminate between the cosmologically interesting regions in the mSUGRA parameter space, and thus obtaining important information about the nature of the annihilating DM.
The same figure also indicates the comparably small contribution from line signals and
states the ratio of IB over secondary photons at high energies. The latter can be sizable especially in the $\tilde\tau$-coannihilation region, thereby significantly improving the detectability of such DM candidates in  future gamma-ray experiments\cite{IB_enhancement}.

 As a further example, the right panel of Fig.~\ref{fig_spectra} compares the spectra of annihilating Wino DM \cite{Bergstrom:2005ss} and Kaluza-Klein DM in theories of universal extra dimensions\cite{Bergstrom:2004cy}; to distinguish these two cases, an energy resolution of $15\%$ is clearly sufficient. Yet another example of the discrimination power of IB-dominated signals is given in Ref.~\refcite{Perelstein:2010at} for DM models that have been invoked to explain the already mentioned anomalies\cite{cosmicray_DM}  in cosmic rays.

\section{Dark matter sources}

When looking for suitable targets for indirect DM searches in gamma rays, one obviously will prefer places with a high (dark) matter density. At least as important, however, is  a low astrophysical background or, rather, a high  signal to background ratio. Since it is quite difficult to constrain the DM distribution from kinematic observations -- especially in the innermost part of the Milky Way, which is \emph{not} DM-dominated -- one usually has to infer the DM density profile from $N$-body simulations. For non-interacting cold DM (like WIMPs), these consistently find cuspy DM distributions in virialized halos;  the most recent, highest-resolution simulations\cite{nbody} tend to somewhat favor the Einasto profile\cite{einasto}
\be
 \rho_\chi(r)=\rho_s\exp\left[-\frac{2}{\alpha}\left(\left(\frac{r}{r_s}\right)^\alpha-1\right)\right]\,,
\ee
at the expense of introducing one more free parameter, over the slightly steeper NFW profile\cite{Navarro:1995iw} which simply scales like $\rho_\chi\propto r^{-1}$ at small distances $r$ from the center. Such simulations, however, have to be taken with a grain of salt since they do not (fully) include the effect of baryons; another  point to keep in mind is that one often has to extrapolate the above scaling behavior several orders of magnitude beyond the actual numerical resolution in order to make quantitative predictions for the expected annihilation flux.

Observations of external galaxies, especially those small in mass and dominated by DM, are sometimes quoted as evidence for shallow DM profiles with a central core\cite{dwarf_cores_obs}. While this evidence is not necessarily conclusive given that at least dwarf galaxies are also consistent with NFW profiles \cite{Strigari:2010un},
such a potential discrepancy with the results from $N$-body simulations could be attributed  to the effect of baryons when including dynamical friction and stellar feedback processes.
\cite{dwarf_cores_theo}.
For \emph{baryon-dominated} systems like our own galaxy, the presence of baryons could actually also lead to a \emph{steepening} of the DM profile in a process known as \emph{adiabatic contraction}\cite{adiab_contr}. This is an issue of ongoing debate and there is evidence both in favour of\cite{adiab_contr_pro} and against\cite{adiab_contr_con} such an additional steepening of the DM profile (for a discussion, see Ref.~\refcite{adiab_contr_disc}).    
The infall of (dark) matter onto a black hole at the center of the halo, like the supermassive black hole at the center of the Milky Way, could lead to an even more spiky DM profile\cite{Colafrancesco:2006he}, though this  strongly depends on the halo formation history\cite{noBHspike}.
The upshot is that the NFW (or Einasto) profile is a natural assumption for WIMP DM -- as long as one keeps in mind that the currently not very well understood role of baryons may somewhat change this picture.

Given this assumption, the \emph{galactic center} is the single-most brightest source of DM annihilation in the sky. At the same time, however, it is also an astrophysically very rich and complex environment, which inevitably means that  it will generally be quite difficult to unambiguously distentangle a DM signal from the background \cite{Zaharijas:2006qb}. The situation becomes further complicated by the fact that  adopting other DM profiles would change the expected gamma-ray flux by several orders of magnitude\cite{Fornengo:2004kj}. 

For that reason, it may be advantageous to broaden the view and include the diffuse gamma-ray emission from the whole Milky Way halo, in which case the expected \emph{angular distribution} of the DM signal would be a nice signature to look for (especially when combined with the observation of spectral signatures)\cite{Dodelson:2007gd}. Indeed, when focussing on the signal to background ratio, there are indications that the optimal angular window of observation is an annulus of a few degrees around the galactic center\cite{Serpico:2008ga}.
There is also a DM-induced contribution to the extra-galactic flux, stemming from annihilations in external halos at higher redshift and then integrated over the cosmological evolution\cite{Ullio:2002pj}. While the challenge lies in the large extragalactic background, which is even less understood than its galactic counterpart, there could be prominent spectral signatures to look for, like the $\pi^0$ peak or line signals which would be modified in a characteristic way due to the cosmological redshift.

As for discrete sources, \emph{dwarf galaxies} are the most DM dominated objects observed so far, with mass-to-light ratios of up to $\sim\!1000$, and therefore represent in some sense the opposite extreme of the galactic center: while the DM-signal in gamma rays is expected to be rather faint, it should by far outshine any astrophysical backgrounds. Consequently, these objects have a long history as targets for DM searches\cite{dwarfs1}, which has been triggered anew\cite{dwarfprofiles,dwarfs2} with the detection\cite{Belokurov:2006ph} of a new class of ultra-faint galaxies. Kinematic observations of the stars in these galaxies allow an ever better determination of their density profile and the resulting uncertainty in the annihilation flux is considerably less than for the galactic center\cite{dwarfprofiles}. 

Numerical simulations confirm the theoretical expectation that DM is not distributed smoothly in the galactic halos but forms a lot of clumps and \emph{substructures}. Large DM clumps, if not massive enough to trigger star formation and thus act as a possible seed for a dwarf galaxy\cite{Strigari:2007ma},  would thus be ideal targets for the purpose of DM detection\cite{clump_target}; if discovered as unidentified gamma-ray sources in all-sky surveys, a follow-up observation with the much better sensitivity of ground-based telescopes  could verify their dark nature. With reasonably optimistic assumptions about the particle nature of the DM, the results of simulations seem to indicate that the discovery of at least a few DM clumps could soon be within reach\cite{clumps_soon}. The overall signal from very small, unresolved clumps is, in fact, already constrained by observations of the diffuse background\cite{Pieri:2007ir}. 

Another interesting target for indirect searches are galaxy clusters which are very massive, DM dominated systems; the gamma-ray background from cosmic rays may, however, complicate the discrimination of an annihilation signal\cite{clusters} and a multi-wavelength oriented approach could therefore be more appropriate\cite{cluster_MW}. Finally, an interesting possibility might be the existence of DM mini-spikes around intermediate mass black holes\cite{Bertone:2005xz} which, however, already seems to be in some tension with current observational data\cite{Bringmann:2009ip}.

The annihilation signal in all these cases could be enhanced\cite{Silk:1992bh} by the existence of halo sub-(sub-)structure which is predicted to extend down to subhalo mass scales of $M_{\rm cut}\sim10^{-11}-10^{-3}M_\odot$\cite{Bringmann:2009vf}, where the numerical value of the cutoff depends on the DM particle properties and is set by the kinetic decoupling of DM in the early universe\cite{Green:2005fa}. 
The reason for this is simply that the line-of sight integral in Eq.~(\ref{flux}) effectively constitutes an average  and one always has $\langle\rho^2\rangle>\langle\rho\rangle^2$ for inhomogeneous distributions. Roughly speaking, the boost factor receives about the same contribution from each logarithmic interval in the subhalo mass, though the usual cautionary remarks on the underlying extrapolation of results from numerical simulations apply here as well. Near the galactic center, the effect should be rather small since most clumps have likely been destroyed due to tidal disruption. One should note that the term "boost-factor" is not used uniquely in the literature and in general depends both on the particle species and the energy (at least for charged particles)\cite{non_universal_bf};  sometimes, it is even used to include particle-physics effects like the recently much discussed Sommerfeld enhancement\cite{sommerfeld}.

\section{Experiments and observational status}

Gamma-ray observations can be performed either in space or on earth. The 
sensitivity of space-based telescopes is generally limited by rather small effective areas 
and there is an upper bound on the photon energy that can be resolved 
($\sim300\,$GeV for the already operating Fermi\cite{fermi} satellite, $\sim1\,$TeV for AMS-02\cite{ams02} which will be installed on the international space station in early 2011). A great field  of view, on the other hand, makes the discovery of previously unknown sources a relatively straight-forward task. Earth-based telescopes like HESS\cite{hess}, VERITAS\cite{veritas}, MAGIC\cite{magic} or the planned AGIS\cite{agis} and Air Cherenkov Telescope Array (CTA)\cite{cta} are complementary in that they 
have to face a lower threshold for  the observable energy since the discrimination of electromagnetic from hadronic showers becomes increasingly difficult and below about 10 GeV
electromagnetic showers produced in the atmosphere do not reach the ground anymore; due to an effective 
area of up to several km$^2$ they can, however, achieve much better sensitivities than 
space-based telescopes. Their rather small feld of view makes them ideal for 
pointed observations. 

A DM induced gamma-ray signal from the galactic center has been claimed almost every time new observational data from this region became available\cite{gcgamma,Bergstrom:2004cy}, but so far more refined analyses and new data always tended to disfavor the DM hypotheses previously put forward\cite{nodm}. The ongoing data taking and analysis of the Fermi collaboration will in any case be a crucial step in determining the background sufficiently well so as to eventually be able to discriminate even a subdominant DM component. In passing, we note that in particular the contribution of light DM particles  to the observed flux is severely bound by observations in other cosmic rays species\cite{nolightDM}.

In the recent past, there have been a number of null searches for DM signals which place limits on the relevant DM properties (its mass and annihilation cross section, in particular) that  start to actually touch the theoretically favored parameter space of standard WIMP candidates. Among these new limits many result from the first year of Fermi data (which could, consequently, be improved considerably during the remaining years of operation) -- including an all-sky search for line-signals\cite{Abdo:2010nc}, cosmological DM annihilations\cite{Abdo:2010dk} in the extragalactic background and detailed observations of galaxy clusters\cite{Ackermann:2010rg} as well as dwarf galaxies\cite{Abdo:2010ex}; an analysis of unidentified point sources, among which there could be DM clumps, is underway.
The major operating Air Cherenkov Telescopes have all identified dwarf galaxies as interesting targets, too\cite{ACTdwarfs}; given the relatively short observation times of only 20-50 h, however, the resulting limits on the annihilation cross section can't compete with the ones derived in Ref.~\refcite{Abdo:2010ex}. Taken at face value, the latter actually represent the overall strongest of such limits; in fact, these results will likely improve even further by a "stacking" of all the dwarf galaxies, including Segue 1 for which the analysis is still ongoing,  i.e.~by using  the fact that the DM spectrum should be the same in each case and that the luminosity \emph{ratios} of these objects are much better constrained than the individual flux from any given dwarf\cite{Strigari:2006rd}. One should of course keep in mind, however, that a comparison of limits derived from different sources is always a delicate issue -- both since the underlying assumptions are often difficult to compare and due to the already discussed, large astrophysical uncertainties involved.

\section{Future prospects and comparison with direct searches}

\begin{figure}[t]
\begin{minipage}[t]{0.48\textwidth}
\centering
\includegraphics[width=\textwidth]{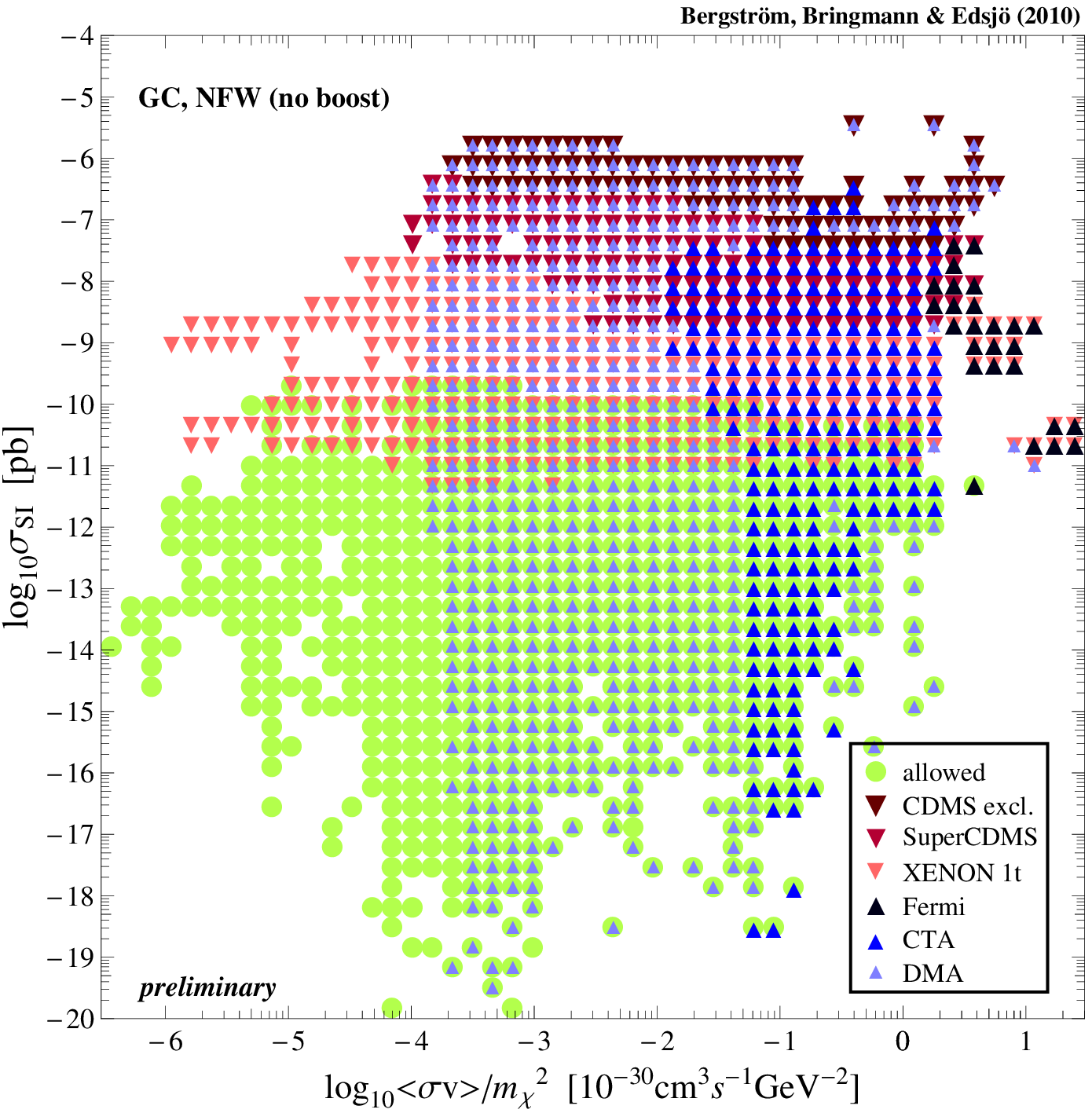}\\
\end{minipage}
\begin{minipage}[t]{0.02\textwidth}
\end{minipage}
\begin{minipage}[t]{0.48\textwidth}
\centering  
\includegraphics[width=\textwidth]{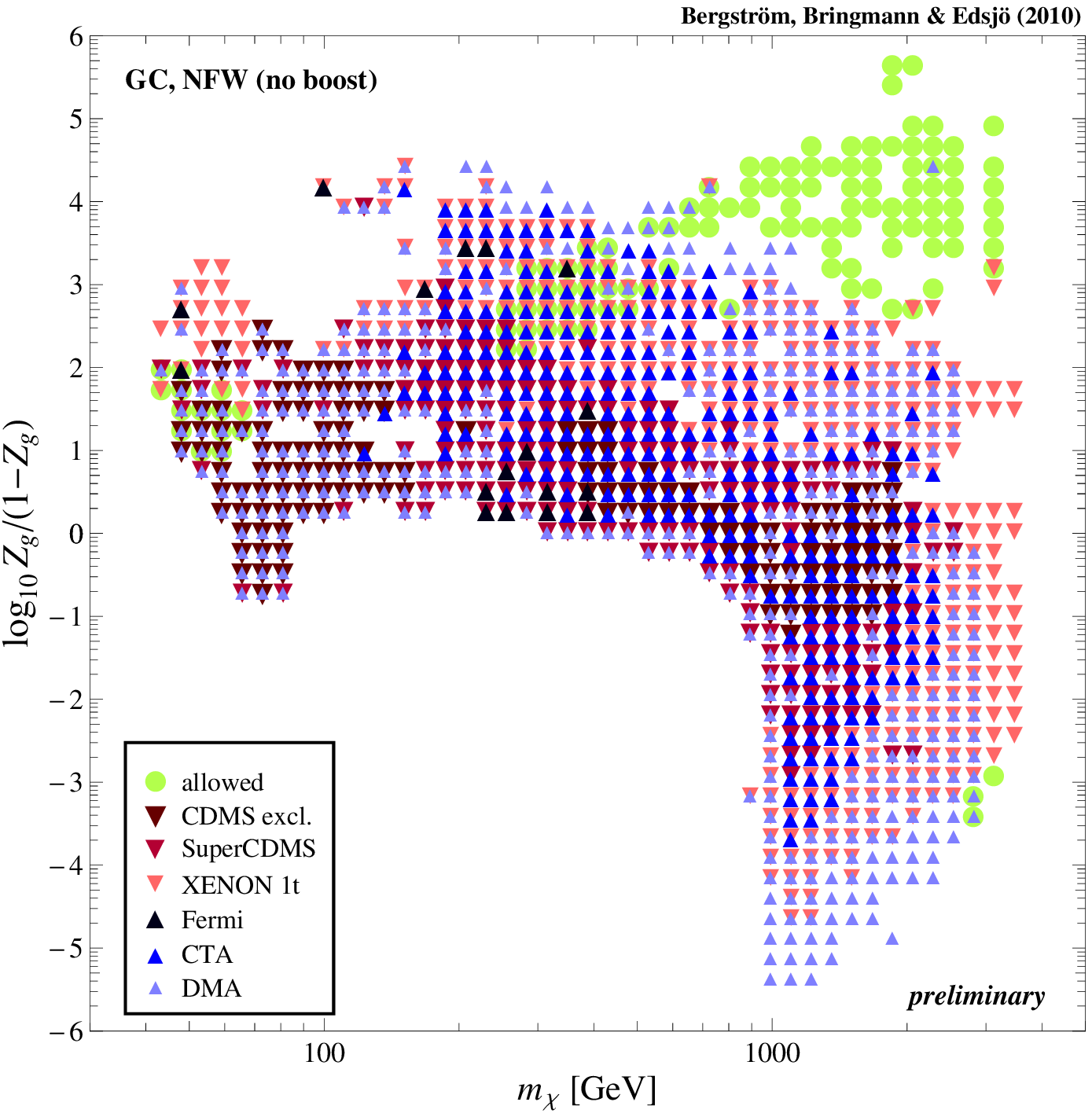}\\
\end{minipage}
\caption{
The reach of direct and indirect detection experiments is shown for the case of neutralino DM, both for presently operating and for future experiments. For this purpose,the left panel shows the spin-independent cross section relevant for direct searches versus a quantity directly relevant for indirect searches, $\langle\sigma v\rangle/m_\chi^2$.
The left panel shows the gaugino fraction $Z_g$ of the neutralino versus its mass $m_\chi$.
Further details are provided in the text; for even more, including an updated version of the figures, see Ref.~\refcite{dirvsindir}.
\label{fig_dirvsindir}}
\end{figure}

As already alluded to, the effective area will always be a limiting factor for space-based experiments. Ground-based detectors, on the other hand, are multi-purpose experiments dedicated to the physics of extreme objects, with DM often considered a mere side-aspect of the overall science goals; even for CTA one can probably not realistically expect to allocate more than about 5\% of the available observation time for DM searches.

In order to assess the real potential of indirect searches, it is therefore timely to think about the possibility  of a \emph{dedicated} DM experiment\cite{dirvsindir} -- a concept that is, indeed, very familiar from direct searches but has not been explored so far for {indirect} searches. To see how far one could possibly get with such an approach, Ref.~\refcite{dirvsindir} introduces the idea of a "Dark Matter Array" (DMA) with a CTA-like setup, but optimized for DM searches. 
Compared to CTA, DMA could gain roughly two orders of magnitude in sensitivity by allowing for an energy threshold of 10 GeV (note that at high altitudes above sea-level, even  5 GeV seems to be possible with current technology\cite{5at5}), an effective area of about 10 km$^2$ and  $5000\,$h observation time instead of the usually considered $50\,$h, implementing thus the idea of a dedicated experiment that only would observe a selected, very small number of targets highly relevant for DM. 

In Fig.~\ref{fig_dirvsindir}, the reach of indirect searches in gamma rays is compared to current and planned direct detection experiments like (Super)CDMS and XENON\cite{dd_exp}. For the indirect searches, the background model\cite{GC_bg} of the Fermi LAT group was employed and it was assumed that the angular resolution is good enough to distinguish the contribution from the galactic center point source\cite{GC_HESS} observed by HESS. The figure shows the result of large scans over the parameter space for the supersymmetric neutralino using \ds\cite{ds}, where only models  were included that satisfy all current collider bounds and give the correct relic density.
Clearly, the parameter space of well-motivated DM candidates extends to much smaller direct detection cross sections than what is usually shown in corresponding exclusion plots; such models would only be accessible by indirect searches and the DMA would be ideal for such a task. Similar conclusions can be drawn for the observation of dwarf galaxies\cite{dirvsindir}; for the galactic center, observational prospects of course improve significantly for clumpy or more spiky profiles than NFW. From the right panel of Fig.~\ref{fig_dirvsindir}, one can see that direct searches are particularly well suited for mixed neutralinos, while very massive Higgsinos could only be probed by a setup like DMA; very massive Gauginos are more difficult to probe and would require rather favorable astrophysical conditions. Finally, it is worth stressing that the most massive models in reach of XENON1t and/or DMA would be out of reach for the LHC, demonstrating once again the complementarity of all these approaches.

\section{Conclusions}

While we have (probably) not seen a DM signal so far, such a detection could well be just around the corner: indirect searches place ever more stringent constraints that start to touch the parameter region of interest for WIMPs -- which one can argue to be the \emph{a priori} best motivated DM candidates  from the point of view of particle physics. Gamma rays are particularly interesting in this respect since they carry \emph{spectral information} which could be used to infer detailed  properties of the particle nature of DM, especially if the observed signal extends to the highest kinematically accessible energies. If one could correlate such results with findings from indirect searches at other wavelengths or from other messengers, this would of course allow an even better determination of the DM properties.

However, one should keep in mind that the parameter space of well-motivated DM candidates extends well beyond the reach of any currently planned indirect detection experiment.
 In order to fully make use of the potential of indirect searches, one would need a \emph{dedicated DM experiment} like the Dark Matter Array (DMA) which would prove truly complementary to other approaches and could, even without very optimistic assumptions about the astrophysical distribution of DM, constrain models that are completely out of reach for direct detection experiments or the LHC.

\section*{Acknowledgments}

I would like to thank the organizers for inviting me to this conference. As an Emmy Noether fellow, I gratefully acknowledge support from the German Research Foundation (DFG) under grant BR 3954/1-1.

\end{document}